\documentclass[twocolumn,showpacs,prd,floatfix,axodraw]{revtex4}
\usepackage{mathrsfs}
\usepackage{graphicx,,booktabs,bm}
\usepackage{overpic}
\usepackage{color}
\usepackage{amssymb}

\usepackage{mathrsfs,bm,amsmath,amssymb}
\usepackage{longtable,lscape}
\usepackage{txfonts}
\usepackage{amssymb}
\usepackage{indentfirst}
\usepackage{graphicx,,booktabs}
\usepackage{multirow}
\usepackage{color}
\usepackage{amssymb}

\definecolor{cover}{rgb}{0.77,0.87,0.88}
\definecolor{blueone}{rgb}{0.1,0.1,.7}
\definecolor{citec}{rgb}{0.14,0.47,0.09}
\definecolor{two}{rgb}{0.0,0.5,0.}
\definecolor{three}{rgb}{.5,.1,0.15}
\usepackage[bookmarks=true,bookmarksopen=false,plainpages=false,breaklinks=true,
   bookmarksnumbered=true,hypertexnames=false,
   filecolor=blue,urlcolor=three,menucolor=three,
   linkcolor=three,citecolor=blueone, colorlinks,
   anchorcolor=blue,runcolor=pink,frenchlinks=red
   pdfstartview=FitH,pdftitle=title,%
   pdfauthor=author]{hyperref}

\begin{document}
\title{{Molecular states from $\bar{B}^{(*)}N$ interactions}}

\author{Zhong-Yi Jian$^{1*}$}
\author{Hong Qiang Zhu$^{2}$ \footnote{These authors contributed equally to this work.} \footnote{corresponding author}} \email{20132013@cqnu.edu.cn}
\author{Feng Yang$^{1}$}
\author{Qi-Hui Chen$^{1}$ \footnote{corresponding author}}\email{qhchen@swjtu.edu.cn}

\author{Yin Huang$^{3,1}$\footnote{corresponding author}}\email{yin.huang@apctp.org}

\author{Jun He$^{4}$\footnote{corresponding author}}\email{junhe@njnu.edu.cn}
\affiliation{$^1$ School of Physical Science and Technology, Southwest Jiaotong University, Chengdu 610031,China}
\affiliation{$^2$College of Physics and Electronic Engineering, Chongqing Normal University, Chongqing 401331,China}
\affiliation{$^3$Asia Pacific Center for Theoretical Physics,Pohang University of Science and Technology, Pohang 37673, Gyeongsangbuk-do,South Korea}
\affiliation{$^4$School of Physics and Technology, Nanjing Normal University, Nanjing 210097, China}

\date{\today}
\begin{abstract}

In 2019, two new structures $\Lambda_b(6146)$ and $\Lambda_b(6152)$ were observed by the LHCb Collaboration at the invariant mass
spectrum of $\Lambda_b^0\pi^{+}\pi^{-}$, which aroused a hot discussion about their inner structures.
The $\Lambda_b(6146)$ and $\Lambda_b(6152)$ might still be molecular states because their masses are close to threshold of a $\bar{B}$ meson and a nucleon.  In this work, we perform a systematical investigation of possible heavy baryonic molecular
states from the  $\bar{B}N$ interaction.  Since the $\bar{B}N$ channel strongly couples to the
$\bar{B}^{*}N$ channel, the possible $\bar{B}N-\bar{B}^{*}N$ bound states are also studied.  The interaction of the system
considered is described by the $t$-channel $\sigma$, $\pi$, $\eta$ ,$\omega$, and $\rho$ mesons exchanges.  By solving the
non-relativistic Schr\"{o}dinger equation with the obtained one-boson-exchange potentials, the $\bar{B}^{(*)}N$ bound states with
different quantum numbers  are searched.  The calculation suggests that recently observed $\Lambda_b(6146)$ can
 be assigned as a $P$-wave $\bar{B}N$  molecular state with spin parity $J^P=3/2^{+}$ or a $\bar{B}N-\bar{B}^{*}N$
bound state.  However, assignment of $\Lambda_b(6152)$ as an $F$-wave $\bar{B}N$ molecular  is disfavored.  The $\Lambda_b(6152)$ can
be explained as meson-baryon molecular state with a small $\bar{B}N$ component.  The calculation also predict the existence of two
$S$-wave $\bar{B}N-\bar{B}^{*}N$ bound states that can be related to the experimentally observed $\Lambda_b(5912)$ and $\Lambda_b(5920)$.

\end{abstract}


\maketitle
\section{INTRODUCTION}
Thanks to the experimental progress, there is a great achievement in searching for exotic states beyond the conventional
quark model~\cite{Zyla:2020zbs}.   Understanding their internal structure and using it as a probe to the production mechanism
is among the most active research fields in hadrons physics.  The molecular
picture is widely adopted and successful to explain or predict the candidates of exotic states.  For example, the best candidates
for molecules are the several narrow hidden-charm pentaquarks $P_{c}$~\cite{LHCb:2019kea} and $P_{cs}$~\cite{LHCb:2020jpq},
which are widely believed to have dominant $\Sigma_c^{(*)}D^{(*)}$ and $\Xi_c\bar{D}^{*}$ components~\cite{Xiao:2019mvs,Yang:2021pio,He:2019rva,Paryev:2022zdx},
respectively.  And the existence of hidden-bottom pentaquark molecular states has been studied in many works~\cite{Wang:2020bjt, Zhu:2020vto}.

The successful application of the molecular picture naturally leads us to study whether the $\bar{B}$ meson and nulceon can be bound
together to form baryonic molecular state corresponding to a $\Lambda_b$ baryon.   At present, there are five $\Lambda_b$
baryons that can be well established with three-star ratings listed in the Review of Particle Physics (PDG)~\cite{Zyla:2020zbs}.  The ground-state
$\Lambda_b$ baryon can be fitted into the conventional quark model.  There exist many different interpretations for the
states $\Lambda_b(5912)$ and $\Lambda_b(5920)$.  In addition to the traditional three quark
state~\cite{Wang:2017kfr, Gandhi:2020otj, Kawakami:2019hpp, Kawakami:2018olq}, they were explained as molecular state containing
the main component of $\bar{B}N$ and $\bar{B}^{*}N$~\cite{Romanets:2013cqa,Lu:2014ina,Liang:2014eba,GarciaRecio:2012db,Huang:2021ave},
respectively.  In other words, whether these states are compact multiquarks or meson-baryon molecules is a matter of intense debate.

In 2019, the other two bottom baryons $\Lambda_b(6146)$ and $\Lambda_b(6152)$ were first observed  by the LHCb Collaboration
in the $\Lambda_b^0\pi^{+}\pi^{-}$ invariant mass spectrum~\cite{LHCb:2019soc}.  The observed resonance masses, widths, isospin (I),
and spin-parity ($J^P$) are, respectively,
\begin{align}
\Lambda_b(6146): M&=6146.17\pm 0.33 \pm 0.22 \pm 0.16~{\rm MeV},\nonumber\\
\Gamma&=2.9\pm{}1.3 \pm 0.3 {\rm MeV},~I(J^P)=0(3/2^{+}),\nonumber\\
\Lambda_b(6152): M&=6152.51\pm 0.26 \pm 0.22 \pm 0.16~~{\rm MeV},\nonumber\\
\Gamma&=2.1\pm{}0.8 \pm 0.3~ {\rm MeV},~~I(J^P)=0(5/2^{+}).\nonumber
\end{align}
The signals consistent with the $\Lambda_b(6146)$ and $\Lambda_b(6152)$ states are confirmed by the CMS Collaboration several
months later~\cite{CMS:2020zzv}.  The $3$-quark structure of these two states were studied by different approaches in the literature~\cite{Mao:2020jln,Azizi:2020tgh,Chen:2019ywy,Wang:2019uaj,Liang:2019aag,Yu:2021zvl,Kakadiya:2021jtv}.

Since the mass of $\Lambda_b(6146)$ and $\Lambda_b(6152)$ are about 65 MeV and 71 MeV below the threshold of $\bar{B}N$
($m_{\bar{B}}=5279.34\pm{}0.12$ MeV and $m_{p}=938.27$ MeV~\cite{Zyla:2020zbs}), respectively, it is novel but plausible to explain the newly observed $\Lambda_b(6146)$ and $\Lambda_b(6152)$ as $\bar{B}N$ bound state. An binding energy in
the range of 5- 100 MeV is one of the simple qualitative signatures for an $S$-wave hadron-pair molecule~\cite{Barnes:1994pd}.
However, the quantum numbers of $\bar{B}$ meson and nucleon are $J^P=0^{-}$ and $1/2^{+}$, respectively. To form a bound state with
the quantum number $J^P=3/2^{+}$ ($J^P=5/2^{+}$), the coupling between $\Lambda_b(6146)$ ($\Lambda_b(6152)$) and its
constituents $\bar{B}N$ should be a $P$ ($F$) wave.  We also find in
Refs.~\cite{Romanets:2013cqa,Lu:2014ina,Liang:2014eba,GarciaRecio:2012db,Huang:2021ave,Carames:2017pfl}
that the $\bar{B}N$ channel strongly couples to the $\bar{B}^{*}N$ channel.  Hence, the coupled-channel interaction of these two
channels is considered.  The study of whether they can form high wave molecular states is very
helpful to understand the interaction between the quarks.

In this work, we study the $\bar{B}^{(*)}N$ interaction within the one-boson exchange (OBE) model and try to understand the
baryons $\Lambda_b(6146)$ and $\Lambda_b(6152)$, which have masses close to the $\bar{B}N$ threshold.
This paper is organized as follows. In Sec.~\ref{Sec: formulism}, we will present the theoretical formalism.
In Sec.~\ref{Sec: results},  the numerical result will be given, followed by discussions and conclusions in the last section.

\section{THEORETICAL FORMALISM}\label{Sec: formulism}
\begin{figure}[h!]
\begin{center}
\includegraphics[bb=60 20 1050 710, clip, scale=0.50]{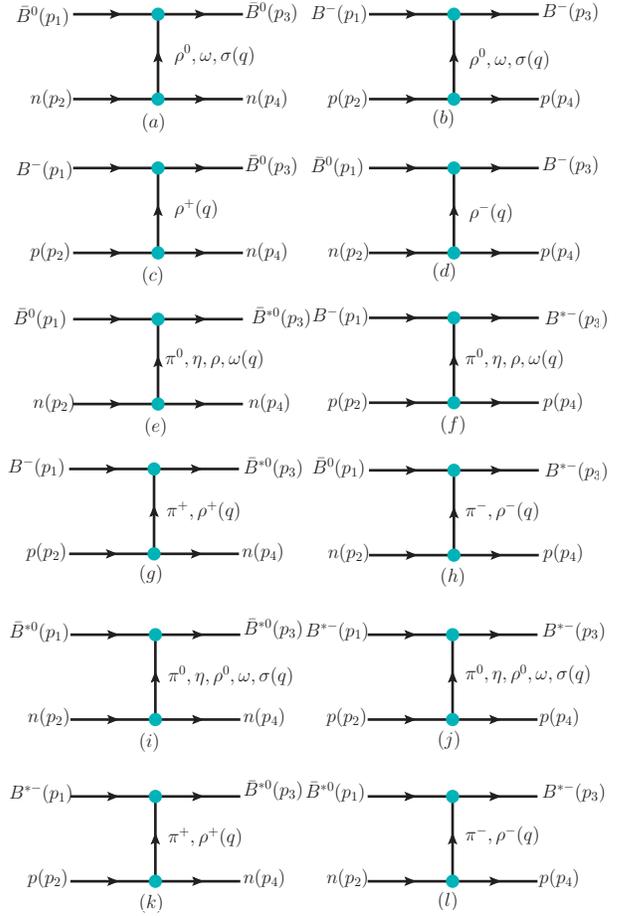}
\caption{Feynman diagram for the processes $\bar{B}^{(*)}N\to{}\bar{B}^{(*)}N$ through vector mesons $\rho,\omega$ and scalar mesons $\pi,\eta,\sigma$ exchanging. }
\label{cc1}
\end{center}
\end{figure}

In this work, we study whether the interaction between $\bar{B}^{(*)}$ meson and nucleon can form a hadronic molecular state corresponding to
a $\Lambda^0_b$ baryon.  The molecular states can be obtained by solving the non-relativistic Schr\"{o}dinger
equation.  To solve the Schr\"{o}dinger equation, the only issue we need to clarify is the explicit form of the $\bar{B}^{(*)}N$
two-body interaction potentials.  In our present work, the $\bar{B}^{(*)}N$ interaction is derived from the OBE model.   And the OBE
model tells us that the simplest Feynman diagram for the $\bar{B}^{(*)}N\to{}\bar{B}^{(*)}N$  process is
the tree diagram, as shown in Fig.~\ref{cc1}.  Since the amplitude for the $\bar{B}^{*}N\to{}\bar{B}N$ reaction is the same as
that the $\bar{B}N\to{}\bar{B}^{*}N$ reaction, only the Feynman diagram for the $\bar{B}N\to{}\bar{B}^{*}N$ processes are shown.
For the system of $\bar{B}^{(*)}N$, the exchanged mesons include vector mesons ${\cal{V}}$, scalar mesons ${\cal{P}}$, and $\sigma$, which read
\begin{align}
{\cal{V}}=
\left(
\begin{array}{ccc}
\frac{\rho^0+\omega}{\sqrt{2}} & \rho^{+}                        & K^{*+}\\
\rho^{-}                       & \frac{-\rho^0+\omega}{\sqrt{2}} & K^{*0}\\
K^{*-}                         & \bar{K}^{*0}                    & \phi  \\
\end{array}
\right),
{\cal{P}}=
\left(
\begin{array}{ccc}
\frac{\sqrt{3}\pi^0+\eta}{\sqrt{6}} & \pi^{+}                              & K^{+}\\
\pi^{-}                             & \frac{-\sqrt{3}\pi^0+\eta}{\sqrt{6}} & K^{0}\\
K^{-}                               & \bar{K}^{0}                          & -\frac{2\eta}{\sqrt{6}}  \\
\end{array}
\right)\nonumber.
\end{align}

To achieve the $\bar{B}^{(*)}N$ interaction potential,  we employ the Lagrangians to depict the vertices of meson exchanges.
Under the chiral symmetry and heavy quark limit, the Lagrangians for heavy mesons interacting with light mesons
read~\cite{Cheng:1992xi, Yan:1992gz, Wise:1992hn, Burdman:1992gh, Casalbuoni:1996pg}
\begin{align}
{\cal{L}}_{\bar{{\cal{B}}}^{*}\bar{{\cal{B}}}^{*}{\cal{P}}}&=-\frac{g}{f_{\pi}}\epsilon_{\alpha\mu\nu\lambda}\bar{{\cal{B}}}^{*\mu\dagger}_{a}\overleftrightarrow{\partial}^{\alpha}\bar{{\cal{B}}}^{*\lambda}_{b}\partial^{\nu}{\cal{P}}_{ba},\\
{\cal{L}}_{\bar{{\cal{B}}}^{*}\bar{{\cal{B}}}{\cal{P}}}&=i\frac{2g\sqrt{m_{\bar{\cal{B}}}m_{\bar{\cal{B}^{*}}}}}{f_{\pi}}(-\bar{{\cal{B}}}^{*\dagger}_{a\lambda}\bar{{\cal{B}}}_{b}+\bar{{\cal{B}}}^{\dagger}_{a}\bar{{\cal{B}}}^{*}_{b\lambda})\partial^{\lambda}{\cal{P}}_{ab},\\
{\cal{L}}_{\bar{{\cal{B}}}\bar{{\cal{B}}}{\cal{V}}}&=-i\frac{\beta{}g_{{\cal{V}}}}{\sqrt{2}}\bar{\cal{B}}^{\dagger}_a\overleftrightarrow{\partial}_{\mu}\bar{\cal{B}}_b{\cal{V}}^{\mu}_{ab},\\
{\cal{L}}_{\bar{\cal{B}}^{*}\bar{\cal{B}}{\cal{V}}}&=\sqrt{2}\lambda{}g_{\cal{V}}\epsilon_{\lambda\alpha\beta\mu}(-\bar{\cal{B}}^{*\mu\dagger}_a\overleftrightarrow{\partial}^{\lambda}\bar{\cal{B}}_b\nonumber\\
                                           &+\bar{\cal{B}}_a^{\dagger}\overleftrightarrow{\partial}^{\lambda}\bar{\cal{B}}^{*\mu}_b)(\partial^{\alpha}{\cal{V}}^{\beta})_{ab},\\
{\cal{L}}_{\bar{\cal{B}}^{*}\bar{\cal{B}}^{*}{\cal{V}}}&=-i\frac{\beta{}g_{\cal{V}}}{\sqrt{2}}\bar{\cal{B}}^{*\dagger}_a\overleftrightarrow{\partial}_{\mu}{\bar\cal{B}}^{*}_b{\cal{V}}^{\mu}_{ab}\nonumber\\
                                               &-i2\sqrt{2}\lambda{}g_{\cal{V}}m_{\bar{\cal{B}}^{*}}\bar{\cal{B}}^{*\mu\dagger}_a\bar{\cal{B}}^{*\nu}_b(\partial_{\mu}{\cal{V}}_{\nu}-\partial_{\nu}{\cal{V}}_{\mu})_{ab},
\end{align}
where $\bar{{\cal{B}}}^{T}=(B^{-},\bar{B}^{0}, \bar{B}_s^{0})$, $A\overleftrightarrow{\partial}^{\lambda}B=A\partial^{\lambda}B-\partial^{\lambda}AB$.
$\epsilon^{\mu\nu\alpha\beta}$ is the Levi-Civit$\grave{a}$ tensor with $\epsilon^{0123}=1$.
The parameters involved here were determined from Refs.~\cite{Falk:1992cx,Isola:2003fh,Liu:2009qhy,Chen:2019asm} and shown in Tabel.~\ref{table1}.
\begin{table}[h!]
\centering
\caption{ The values of the parameters.
}\label{table1}
\begin{tabular}{cccccccc}
\hline\hline
$g_{\cal{V}}=5.9$  ~~~&$g=0.59$    ~~~&$\lambda=0.56$ GeV$^{-1}$    ~~~& $\beta=0.9$      ~~~&$f_{\pi}=132$ MeV   \\
$F=0.51$           ~~~&$D=0.75$    ~~~&$f=93$ MeV                      \\ \hline
\hline
\end{tabular}
\end{table}

The lagrangians for vertices $BB\sigma$, $B^{*}B^{*}\sigma$, and $NN\sigma$ are shown as~\cite{Bardeen:2003kt,Zhao:2014gqa}
\begin{align}
{\cal{L}}_{BB\sigma}&=-2 g_{\sigma{}BB}m_B \bar{B} B \sigma,\\
{\cal{L}}_{B^{*}B^{*}\sigma}&=2 g_{\sigma{}BB}m_{B^{*}} \bar{B}^{*\mu} B^{*}_{\mu} \sigma,\\
{\cal{L}}_{NN\sigma}&=2 g_{\sigma{}NN}m_N \bar{N} \sigma N,
\end{align}
where the coupling constant $g_{\sigma{}BB}=1/3g_{\sigma{}NN}=g_{\pi}/2\sqrt{6}$ with $g_{\pi}=3.73$\cite{Bardeen:2003kt,Zhao:2014gqa}.
The $m_B$ and $m_N$ are the masses of $B$ meson and nucleon, respectively.

Moreover, meson-baryon interactions are also needed and can be obtained from the following chiral Lagrangians~\cite{Garzon:2012np,Oset:1997it}
\begin{align}
{\cal{L}}_{{\cal{V}}bb}&=g(\langle\bar{b}\gamma_{\mu}[{\cal{V}}^{\mu},b]\rangle+\langle\bar{b}\gamma_{\mu}b\rangle\langle{\cal{V}}^{\mu}\rangle)\label{eq7},
\end{align}
\begin{align}
{\cal{L}}_{{\cal{P}}bb}&=\frac{F}{2}\langle\bar{b}\gamma_{\mu}\gamma_5[u^{\mu},b]\rangle+\frac{D}{2}\langle\bar{b}\gamma_{\mu}\gamma_5\{u^{\mu},b\}\rangle,
\end{align}
where the $u^{\mu}=-\sqrt{2}\partial^{\mu}{\cal{P}}/f$ and the parameters $F$ and $D$ can also be found in Tabel.~\ref{table1}.
The $\langle...\rangle$ denotes trace in the flavor space and $b$ is the $SU(3)$ matrix of the baryon octet
\begin{align}
b=
\left(
\begin{array}{ccc}
\frac{\Sigma^0}{\sqrt{2}}+\frac{\Lambda}{\sqrt{6}} & \Sigma^{+}                                          & p\\
\Sigma^{-}                                         & -\frac{\Sigma^0}{\sqrt{2}}+\frac{\Lambda}{\sqrt{6}} & n\\
\Xi^{-}                                            & \Xi^{0}                                             & -\frac{2}{\sqrt{6}}\Lambda  \\
\end{array}
\right).
\end{align}
The coupling constant $g$ in Eq.~(\ref{eq7}) is fixed from the strong decay width of $K^{*}\to{}K\pi$.
With help of the following Lagrangian
\begin{align}
{\cal{L}}_{{\cal{V}}{\cal{P}}\cal{P}}=-ig\langle[{\cal{P}},\partial_{\mu}{\cal{P}}]{\cal{V}}^{\mu}\rangle,
\end{align}
the two body decay width $\Gamma(K^{*+}\to{}K^0\pi^{+})$ is related to $g$ as
\begin{align}
\Gamma(K^{*+}\to{}K^0\pi^{+})=\frac{g^2}{6\pi{}m^2_{K^{*+}}}{\cal{P}}^3_{\pi{}K^{*}}=\frac{2}{3}\Gamma_{K^{*+}},
\end{align}
where the ${\cal{P}}_{\pi{}K^{*}}$ is the three momentum of the $\pi$ meson in the rest frame of the $K^{*}$ meson.
Using the experimental strong decay width, $\Gamma_{K^{*+}}=50.3\pm 0.8$ MeV~\cite{Zyla:2020zbs}, and
the masses of the particles listed in Table.~\ref{table2}, we obtain $g=4.64$.
\begin{table}[h!]
\centering
\caption{Masses of the particles needed in the present work (in units of MeV).
}\label{table2}
\begin{tabular}{ccccccccc}
\hline\hline
~$\pi^0$    ~&$\sigma$       ~&$\pi^{\pm}$   ~&$K^0$          ~& $K^{\pm}$      ~&$K^{*0}$   ~&$K^{*\pm}$     ~ \\
~134.977    ~&550.00         ~&139.57        ~&497.611        ~& 493.68         ~&895.55     ~&891.76         ~ \\
~$B^{-}$    ~&$\eta$         ~& $B^{*}$      ~&$p$            ~& $\rho^0$       ~&$\omega$   ~&$\bar{B}^{0}$  ~ \\
~5279.34    ~&547.86         ~& 5324.70      ~&938.27         ~& 775.26         ~& 782.66    ~&5279.65        ~ \\ \hline
\hline
\end{tabular}
\end{table}

Because hadrons are not pointlike particles, we need to include the form factors in evaluating the scattering
amplitudes of the $\bar{B}^{(*)}N\to{}\bar{B}^{(*)}N$ reaction.  For the $t$-channel $\rho$, $\omega$, $\eta$,
and $\sigma$ mesons exchange, we would like to apply a widely used pole form factor, which is
\begin{align}
{\cal{F}}_{i}=\frac{\Lambda_i^2-m_i^2}{\Lambda_i^2-q_i^2},~~~~~~~ i=\rho,\omega,\sigma,\eta,
\end{align}
where $m_i$ and $q_i$ are the masses and the four-momenta of exchanged mesons, respectively. $\Lambda_i$ is a
cutoff and will be taken as a parameter and discussed later.

With the above prepared, we can obtain the general expressions of the amplitudes corresponding to the Feynman diagrams Fig.~\ref{cc1}
\begin{align}
{\cal{M}}_{a/b}&=-i\frac{\beta g g_v}{2\sqrt{2}}\bar{u}(p_4,s_4)\gamma^{\theta}u(p_2,s_2)\frac{-g^{\theta\mu}+q^{\theta}q^{\mu}/m_{\rho^0}^2}{q^2-m_{\rho^0}^2}\nonumber\\
             &\times{}(p_3^{\mu}+p_1^{\mu}){\cal{F}}^2_{\rho^0}-i\frac{3\beta{}gg_v}{2\sqrt{2}}\bar{u}(p_4,s_4)\gamma^{\theta}u(p_2,s_2)\nonumber\\
             &\times\frac{-g^{\theta\mu}+q^{\theta}q^{\mu}/m_{\omega}^2}{q^2-m_{\omega}^2}(p_3^{\mu}+p_1^{\mu}){\cal{F}}^2_{\omega}-i4g_{\sigma{}BB}g_{\sigma NN }m_Nm_B\nonumber\\
              &\times\bar{u}(p_4,s_4)u(p_2,s_2)\frac{1}{q^2-m_{\sigma}^2}{\cal{F}}^2_{\sigma}\label{eq10},
\end{align}
\begin{align}
{\cal{M}}_{c/d}&=-i\frac{\beta{}gg_v}{\sqrt{2}}\bar{u}(p_4,s_4)\gamma^{\theta}u(p_2,s_2)\frac{-g^{\theta \mu }+q^{\theta}q^{\mu}/m_{\rho^{\pm}}^2}{q^2 - m_{\rho^{\pm}}^2}\nonumber\\
             &\times(p_3^{\mu}+p_1^{\mu}){\cal{F}}^2_{\rho^{\pm}}\label{eq11},\\
{\cal{M}}_{e/f}&= -\frac{g \sqrt{m_{B} m_{B^*}}}{\sqrt{2} f f_{\pi} }[\frac{D+F}{q^2 - m_{\pi^0}^2}{\cal{F}}^2_{\pi^0}-\frac{D-3F}{3(q^2-m_{\eta}^2)}{\cal{F}}^2_{\eta^0}]\nonumber\\
              &\times{}q\cdot \epsilon(p_1,s_1)\bar{u}(p_4,s_4)q\!\!\!/ \gamma^5 u(p_2,s_2)\nonumber\\
              &+i\frac{\lambda}{ \sqrt{2} } g g_v \epsilon_{\lambda \alpha \beta \mu } \bar{u}(p_4,s_4) \gamma^{\theta}u(p_2,s_2)[\frac{-g^{\theta
               \beta}+q^{\theta}q^{\beta}/m_{\rho^0}^2}{q^2 -m_{\rho^0}^2}{\cal{F}}^2_{\rho^0}\nonumber\\
               &+3\frac{-g^{\theta \beta }+q^{\theta}q^{\beta}/m_{\omega}^2}{q^2 - m_{\omega}^2}{\cal{F}}^2_{\omega}](p_3^{\lambda}+p_1^{\lambda})q^{\alpha} \epsilon^{\mu}(p_1,s_1),\\
{\cal{M}}_{g/h}&= \frac{\sqrt{2}g\sqrt{m_{\bar{B}}m_{\bar{B}^{*}}}(D+F)}{ff_{\pi} }\bar{\mu}(p_4,s_4)q\!\!\!/\gamma^5\mu(p_2,s_2)\frac{1}{q^2 - m_{\pi^{\pm}}^2}\nonumber\\
               &\times{}q^{\lambda}\epsilon^{\lambda\dagger}(p_3,s_3){\cal{F}}^2_{\pi^{\pm}}+i \sqrt{2}\lambda{}g g_v \epsilon_{\lambda\alpha\beta\mu}\bar{\mu}(p_4,s_4)\gamma^{\theta}\mu(p_2,s_2)\nonumber\\
               &\times\frac{-g^{\theta\beta}+q^{\theta}q^{\beta}/m_{\rho^+}^2}{q^2- m_{\rho^{\pm}}^2}(p_3^{\lambda}+p_1^{\lambda})q^{\alpha}\epsilon^{\mu\dagger}(p_3,s_3){\cal{F}}^2_{\rho^{\pm}},\\
{\cal{M}}_{i/j}&=-i\frac{g}{2\sqrt{2}ff_{\pi}}\epsilon_{\alpha\mu\nu\lambda}\bar{\mu}(p_4,s_4)q\!\!\!/\gamma^5\mu(p_2,s_2)(p_3^{\alpha}+p_1^{\alpha})q^{\nu}\nonumber\\
	&\times \epsilon^{\mu}(p_1,s_1)\epsilon^{\dagger\lambda}(p_3,s_3)[\frac{D+F}{q^2 - m_{\pi^0}^2}{\cal{F}}^2_{\pi^0}-\frac{D-3F}{3(q^2-m_{\eta}^2)}{\cal{F}}^2_{\eta^0}]\nonumber\\
    &-\frac{g}{\sqrt{2}}\bar{\mu}(p_4,s_4)\gamma^{\theta}\mu(p_2,s_2)[\frac{-g^{\theta\beta}+q^{\theta}q^{\beta}/m_{\rho^0}^2}{q^2 -m_{\rho^0}^2}{\cal{F}}^2_{\rho^0}\nonumber\\
    &+3\frac{-g^{\theta\beta}+q^{\theta}q^{\beta}/m_{\omega}^2}{q^2-m_{\omega}^2}{\cal{F}}^2_{\omega}]\bigg(\frac{\beta g_v}{2}g^{\alpha\beta}(p_3^{\lambda}+p_1^{\lambda}) +2\lambda{}g_vm_{B^*}\nonumber\\
	&\times(q^{\alpha}g^{\lambda\beta}-q^{\beta}g^{\alpha\lambda})\bigg)\epsilon^{\alpha}(p_1,s_1)\epsilon^{\dagger\beta}(p_3,s_3)+4g_{\sigma}g_{\sigma NN}m_N m_{B^*}\nonumber\\
    &\times{}\bar{\mu}(p_4,s_4)\mu(p_2,s_2)\frac{{\cal{F}}^2_{\sigma}}{q^2-m_{\sigma}^2}\varepsilon(p_1)\cdot\varepsilon^{\dagger}(p_3,s_3),\\
{\cal{M}}_{k/l}&= i\frac{g(D+F)}{\sqrt{2}ff_{\pi}}\epsilon_{\alpha\mu\nu\lambda}\bar{\mu}(p_4,s_4)q\!\!\!/\gamma^5\mu(p_2,s_2)\frac{{\cal{F}}^2_{\pi^{\pm}}}{q^2-m_{\pi^{\pm}}^2}
               (p_3^{\alpha}+p_1^{\alpha}) \nonumber\\
	           &\times{}q^{\nu}\epsilon^{\lambda}(p_1,s_1)\epsilon^{\mu\dagger}(p_3,s_3)+g\bar{\mu}(p_4,s_4)\gamma^{\theta}\mu(p_2,s_2)\nonumber\\
               &\times\frac{-g^{\theta\lambda}
               +q^{\theta}q^{\lambda}/m_{\rho^+}^2}{q^2-m_{\rho^+}^2}\bigg(\frac{\beta g_v}{\sqrt{2}}g^{\alpha\beta}(p_3^{\lambda}+p_1^{\lambda})-2\sqrt{2}\lambda{}g_vm_{B^*}\nonumber\\
               &\times(q^{\alpha}g^{\lambda\beta}-q^{\beta}g^{\alpha\lambda})\bigg)\epsilon^{\alpha}(p_1,s_1)\epsilon^{\beta\dagger}(p_3,s_3)  	
\end{align}
where $(s_4,p_4)$, $(s_3,p_3)$ and $(s_2,p_2)$, $(s_1,p_1)$ denote the spin polarization variables and the four-momenta of the outgoing nucleon, $\bar{B}^{*}$ and the initial nucleon, $\bar{B}^{*}$, respectively.  The polarization vector $\epsilon^{\mu}(s)$ represent the wave function of the spin-1 field and can be expressed as
\begin{align}
\epsilon^{\mu}(s=0)=(0,0,0,-1);~~~\epsilon^{\mu}(s=\pm)=\frac{1}{\sqrt{2}}(0,\pm{}1,i,0).
\end{align}
The $u(p_2,s_2)$ and $\bar{u}(p_4,s_4)$ stand for the spin wave function of nucleon.  In our calculation we adopt the Dirac spinor as
\begin{align}
u(\vec{q},s)&=\sqrt{\frac{E+m}{2m}}
\left(
  \begin{array}{ccc}
    1\\
    \frac{\vec{\sigma}\cdot\vec{q}}{E+m}\\
  \end{array}
\right){\cal{\chi}}_{s},
\end{align}
\begin{align}
\bar{u}(\vec{q},s)&=u^{\dagger}(\vec{q},s)\gamma^0={\cal{\chi}}^{\dagger}_{s}\sqrt{\frac{E+m}{2m}}
\left(
  \begin{array}{ccc}
    1 & -\frac{\vec{\sigma}\cdot\vec{q}}{E+m}\\
  \end{array}
\right),
\end{align}
where ${\cal{\chi}}_{1/2}=(1,0)^{\dagger}$ and ${\cal{\chi}}_{-1/2}=(0,1)^{\dagger}$.  $E$, $\vec{q}$, and $m$
are the energy, three-momenta, and masses of nucleon, respectively, and they have relation $E^2=\vec{q}^2+m^2$.

In the above equations, the initial four-momenta are defined as $p_1=(E_1,\vec{p})$ and $p_2=(E_2,-\vec{p})$,
while the final four-momenta are $p_3=(E_1,\vec{p}^{'})$ and $p_4=(E_2,-\vec{p}^{'})$.  Thus, the four-momentum of the
propagator is
\begin{align}
q=p_3-p_1=p_2-p_4=(0, \vec{p}^{'}-\vec{p})=(0,\vec{q}).
\end{align}

Once the scattering amplitudes obtained, the effective potentials in momentum space can be easy computed by using
the Breit approximation~\cite{Breit:1929zz,Breit:1930zza}
\begin{align}
{\cal{V}}(\vec{q})=-\frac{\mathcal{M}}{\sqrt{\prod_{i}2{m}_{i}\prod_{f}2{m}_f}},
\end{align}
where $m_i$ and $m_f$ are the masses of the initial states and final states, respectively. Since we interpret the
bound states as the eigenvalue of the Schr\"{o}dinger equation, we need to translate the effective
potentials into a form in the coordinates space. In order to obtain the effective potential in  the coordinate space,
the following Fourier transformation is performed
\begin{align}
{\cal{V}}(r)=\int\frac{d^3\vec{q}}{(2\pi)^3}e^{i\vec{q}\cdot{}\vec{r}}{\cal{V}}(\vec{q}).
\end{align}
Their explicit expressions are shown in Appendix.A with the help of the following functions
\begin{align}
{\cal{F}}&\left\{\frac{1}{\vec{q}^2+m^2}(\frac{\Lambda^2-m^2}{\Lambda^2+\vec{q}^2})^2 \right\}\equiv Y_{1}(\Lambda,m,r),\\
{\cal{F}}&\left\{\frac{\vec{q}^2}{\vec{q}^2+m^2}(\frac{\Lambda^2-m^2}{\Lambda^2+\vec{q}^2})^2\right\}=-\nabla_r^2Y_1(\Lambda,m,r),\\
{\cal{F}}&\left\{ \frac{\vec{k}^2}{\vec{q}^2+m^2}(\frac{\Lambda^2-m^2}{\Lambda^2+\vec{q}^2})^2 \right\}=\frac{1}{4}\nabla_r^2Y_1(\Lambda,m,r)-\dfrac{1}{2}{\cal{F}}_1,\\
{\cal{F}}&\left\{ \frac{i\vec{\sigma}\cdot(\vec{q}\times\vec{k})}{\vec{q}^2+m^2}(\frac{\Lambda^2-m^2}{\Lambda^2+\vec{q}^2})^2\right\}
         =\vec{\sigma}\cdot\vec{L}(\frac{1}{r}\frac{\partial}{\partial r})Y_1(\Lambda,m,r)\label{eq18},\\
{\cal{F}}&\left\{ \frac{(\vec{A}\cdot\vec{q})(\vec{B}\cdot \vec{q})}{\vec{q}^2+m^2}(\frac{\Lambda^2-m^2}{\Lambda^2+\vec{q}^2})^2\right\}
         =\frac{1}{3}(\vec{A}\cdot\vec{B})(-\nabla_r^2Y_1(\Lambda,m,r))\nonumber\\
         &+\frac{1}{3}S(\hat{r},\vec{A},\vec{B})(-r\frac{\partial}{\partial r}\frac{1}{r}\frac{\partial}{\partial r}Y_1(\Lambda,m,r)),\\
{\cal{F}}&\left\{\frac{(\vec{A}\cdot\vec{k})(\vec{B}\cdot \vec{q})}{\vec{q}^2+m^2}(\frac{\Lambda^2-m^2}{\Lambda^2+\vec{q}^2})^2\right\}
       =\frac{1}{3}(\vec{A}\cdot\vec{B})(\nabla^2_rY_1(\Lambda,m,r))\nonumber\\
      &-\frac{1}{3}(-r\frac{\partial}{\partial r}\frac{1}{r}\frac{\partial}{\partial r}Y_1(\Lambda,m,r))\nonumber\\
      &-\frac{2}{3}(\frac{\partial}{\partial r}Y_1(\Lambda,m,r))(S(\hat{r},\vec{A},\vec{B})+\vec{A}\cdot\vec{B})\frac{\partial}{\partial{}r},
\end{align}
where ${\cal{F}}$ denotes the Fourier transformation, and $\nabla^2_r=\frac{1}{r^2}\frac{\partial}{\partial{r}}r^2\frac{\partial}{\partial{r}}$.
$\vec{\sigma}\cdot\vec{L}$ is the spin-orbital operator, and $S(\hat{r},\vec{x},\vec{y})=3(\hat{r}\cdot\vec{x})(\hat{r}\cdot\vec{y})-\vec{x}\cdot\vec{y}$
is tensor operator.  Variables in the above functions denote $\vec{q}=\vec{p}_3-\vec{p_1}$ and $\vec{k}=\frac{1}{2}(\vec{p}_3+\vec{p_1})$.
The function $Y_{1}(\Lambda,m,r)$ and recoil correction ${\cal{F}}_1$ are defined as
\begin{align}
Y_{1}(\Lambda,m,r)=\frac{1}{4\pi r}(e^{-m r}-e^{-\Lambda r})-\frac{\Lambda^2-m^2}{8\pi\Lambda}e^{-\Lambda r}
\end{align}
and
\begin{align}
{\cal{F}}_1(m)=\{\nabla_r^2,Y_1(\Lambda,m,r)\}=\nabla_r^2Y_1(\Lambda,m,r) +Y_1(\Lambda,m,r)\nabla_r^2.
\end{align}
Usually, the recoil correction is ignored due to the small contribution.  In this work, we will give a discussion about its role.

Since the isospin of $\Lambda_b^0$ baryons is zero, the flavor wave function $|I, I_3\rangle$ for the $\bar{B}^{(*)}N$ systems are
simple and read as
\begin{align}
|0,0\rangle=\sqrt{1/2}(|B^{(*)-}p\rangle-|\bar{B}^{(*)0}n\rangle),
\end{align}
where $I$ and $I_3$ are the isospin and its third component.  Thus, the total potential in the coordinate space for the discussed systems
are combined as
\begin{align}
{\cal{V}}_{\bar{B}N\to{}\bar{B}N}&=\frac{1}{\sqrt{2}}[V_{a}(r)_{\rho^0,\omega,\sigma}-V_{c}(r)_{\rho^{+}}+V_{b}(r)_{\rho^0,\omega,\sigma}-V_{d}(r)_{\rho^{-}}],\nonumber\\
{\cal{V}}_{\bar{B}^{*}N\to{}\bar{B}^{*}N}&=\frac{1}{\sqrt{2}}[V_{i}(r)_{\pi^0,\eta,\rho^0,\omega,\sigma}-V_{k}(r)_{\rho^{+}}+V_{j}(r)_{\pi^0,\eta,\rho^0,\omega,\sigma}\nonumber\\
         &-V_{l}(r)_{\rho^{-}}-V_{k}(r)_{\pi^{+}}-V_{l}(r)_{\pi^{-}}],\nonumber\\
{\cal{V}}_{\bar{B}^{*}N\to{}\bar{B}N}&={\cal{V}}_{\bar{B}N\to{}\bar{B}^{*}N}=\frac{1}{\sqrt{2}}[V_{e}(r)_{\pi^0,\eta,\rho^0,\omega}+V_{f}(r)_{\pi^0,\eta,\rho^0,\omega}\nonumber\\
                                     &-V_{g}(r)_{\pi^{+},\rho^{+}}-V_{h}(r)_{\pi^{-},\rho^{-}}].
\end{align}

\begin{table}[h!]
\centering
\caption{Possible quantum numbers for $\bar{B}^{(*)}N$ systems involved in our calculation.  The first column contains the spin-parity quantum numbers
corresponding to the channels.  $ A\sim{}B$ stand for mixing effect between $A$ and $B$.}\label{table1-thre}
\begin{tabular}{cccccccc}
\hline\hline
~~~~~~~~~~$J^P$        ~~~~~~~~~~~~&$\bar{B}^{*}N$                                   ~~~~~~~~~~~~& $\bar{B}N$                  ~~~~~~~~~~~~     \\ \hline
~~~~~~~~~~$1/2^{-}$    ~~~~~~~~~~~~& $|^2S_{1/2}\sim^4D_{1/2}\rangle$                ~~~~~~~~~~~~& $|^2S_{1/2}\rangle$         ~~~~~~~~~~~~    \\
~~~~~~~~~~$1/2^{+}$    ~~~~~~~~~~~~& $|^2P_{1/2}\sim^4P_{1/2}\rangle$                ~~~~~~~~~~~~& $|^2P_{1/2}\rangle$         ~~~~~~~~~~~~    \\
~~~~~~~~~~$3/2^{-}$    ~~~~~~~~~~~~& $|^4S_{3/2}\sim^2D_{3/2}\sim^4D_{3/2}\rangle$   ~~~~~~~~~~~~& $|^2D_{3/2}\rangle$         ~~~~~~~~~~~~    \\
~~~~~~~~~~$3/2^{+}$    ~~~~~~~~~~~~& $|^2P_{3/2}\sim^4P_{3/2}\sim^4F_{3/2}\rangle$   ~~~~~~~~~~~~& $|^2P_{3/2}\rangle$         ~~~~~~~~~~~~    \\
~~~~~~~~~~$5/2^{-}$    ~~~~~~~~~~~~& $|^2D_{5/2}\sim^4D_{5/2}\rangle$                ~~~~~~~~~~~~& $|^2D_{5/2}\rangle$         ~~~~~~~~~~~~    \\
~~~~~~~~~~$5/2^{+}$    ~~~~~~~~~~~~& $|^4P_{5/2}\sim^2F_{5/2}\sim^4F_{5/2}\rangle$   ~~~~~~~~~~~~& $|^2F_{5/2}\rangle$         ~~~~~~~~~~~~    \\
\hline
\hline
\end{tabular}
\end{table}

Now we compute the spin-spin interactions and tensor force operators that shown in Table.~\ref{tableqw}.  To perform the calculation,
we should first present the spin-orbital wave functions related to the spin angular momentum $S$, orbit angular momentum $L$, and total
angular momentum $J$ for the discussed system.  The general expressions for the spin-orbital wave functions are written as
\begin{align}
&|\bar{B}N(^{2S+1}L_J)\rangle=\sum_{m_S,m_L}{\cal{C}}_{Sm_S,Lm_L}^{J,M}\chi_{m_s}|Y_{L,m_L}\rangle,\\
&|\bar{B}^{*}N(^{2S+1}L_J)\rangle=\sum^{m_S,m_L}_{m,m^{'}}{\cal{C}}_{1/2m,1m^{'}}^{S,m_s}{\cal{C}}_{Sm_S,Lm_L}^{J,M}\chi_{1/2m}\epsilon^{\mu}_{m^{'}}|Y_{L,m_L}\rangle,
\end{align}
where ${\cal{C}}_{ab,cd}^{e,f}$ is the Clebsch-Gordan coefficient, and $|Y_{L,m_L}\rangle$ is the spherical harmonics function.
$M$, $m_S$, and $m_L$ are the third components of the total angular momentum $J$, the spin angular momentum $S$, and the orbit angular
momentum $L$, respectively.  With the notation $|^{2S+1}L_J\rangle$ assignment, the corresponding spin-orbit wave functions for
the discussed $\bar{B}^{(*)}N$ systems are summarized in Table~\ref{table1-thre}.  Finally, the spin-spin interactions and tensor force operators can be easily
computed by a serial of matrix elements $\langle^{2S^{'}+1}L^{'}_{J^{'}}|{\cal{Z}}|^{2S+1}L_J\rangle$, and the results are
collected in Table.~\ref{tableqw}.  A detail derivation is presented as an example in Ref.~\cite{Yalikun:2021bfm}.
\begin{table*}
	\centering
	\caption{The matrix elements of two-body interaction operators for $\bar{B}^{(*)}N$ systems.}\label{tableqw}
	\begin{tabular}{cccccccc}
		\hline\hline
		$\bar{B}^{(*)}N\to{}\bar{B}^{(*)}N$    &$1/2^{-}$                    & $3/2^{-}$                               &$5/2^{-}$                   &$1/2^{+}$                   &$3/2^{+}$                               &$5/2^{+}$   \\
${\cal{Z}}$		&($^{2}S_{1/2}$,$^{4}D_{1/2}$)& ($^{4}S_{3/2},^{2}D_{3/2},^{4}D_{3/2}$) &($^{2}D_{5/2},^{4}D_{5/2}$) &($^{2}P_{1/2},^{4}P_{1/2}$) &($^{2}P_{3/2},^{4}P_{3/2},^{4}F_{3/2}$) &($^{4}P_{5/2},^{2}F_{5/2},^{4}F_{5/2}$)\\ \hline
		$S(\hat{r},\vec{\sigma},-i\vec{\epsilon}^{\dagger}_{s_3}\times{}\vec{\epsilon}_{s_1})$
		&$\left(\begin{array}{cc}
			0          &  -\sqrt{2}   \\
			-\sqrt{2}   &  -2          \\
		\end{array}\right) $
		& $\left(
		\begin{array}{ccc}
			0  &  1   &   2     \\
			1  &  0   &   -1    \\
			2  &  -1  &   0    \\
		\end{array}
		\right)$
		& $\left(
		\begin{array}{cc}
			0           &  \sqrt{2/7}   \\
			\sqrt{2/7}  &  \frac{10}{7} \\
		\end{array}
		\right)$
		&$\left(
		\begin{array}{cc}
			0          &  -\sqrt{2}   \\
			-\sqrt{2}  &  -2          \\
		\end{array}
		\right)$
		&$\left(
		\begin{array}{ccc}
			0                   &  \frac{1}{\sqrt{5}}  &   -\frac{3}{\sqrt{5}}    \\
			\frac{1}{\sqrt{5}}  &  \frac{8}{5}         &   \frac{6}{5}            \\
			-\frac{3}{\sqrt{5}} &  \frac{6}{5}         &   -\frac{8}{5}           \\
		\end{array}
		\right)$
		&$\left(
		\begin{array}{ccc}
			-\frac{2}{5}        &  \sqrt{\frac{6}{5}}        &   \frac{4\sqrt{6}}{5}     \\
		 \sqrt{\frac{6}{5}}         &  0                   &   -\frac{2}{\sqrt{5}}     \\
			\frac{4\sqrt{6}}{5}  &  -\frac{2}{\sqrt{5}} &   \frac{2}{5}             \\
		\end{array}
		\right)$ \\
		$\vec{\sigma}\cdot(-i\vec{\epsilon}_{s_3}^\dagger\times\vec{\epsilon}_{s_1})$&
		$\left(
		\begin{array}{cc}
			-2        ~~~&   0        \\
			0       ~~~&   1         \\
		\end{array}
		\right)$  &
		$\left(
		\begin{array}{ccc}
			1        &  0         &   0     \\
			0        &  -2        &   0     \\
			0        &  0         &   1     \\
		\end{array}
		\right)$ &
		$\left(
		\begin{array}{cc}
			-2        ~~~&   0        \\
			0       ~~~&   1         \\
		\end{array}
		\right)$&
		$\left(
		\begin{array}{cc}
			-2        ~~~&   0        \\
			0       ~~~&   1         \\
		\end{array}
		\right)$&
		$\left(
		\begin{array}{ccc}
			-2       &  0         &   0     \\
			0        &  1         &   0     \\
			0        &  0         &   1     \\
		\end{array}
		\right)$&
		$\left(
		\begin{array}{ccc}
			1        &  0         &   0     \\
			0        &  -2        &   0     \\
			0        &  0         &   1     \\
		\end{array}
		\right)$
		\\
		$\vec{\sigma}\cdot\vec{\epsilon}_{s_1}$ &
		$(\sqrt{3},0)$ &
		$(0,\sqrt{3},0)$&
		$(\sqrt{3},0)$&
		$(\sqrt{3},0)$&
		$(\sqrt{3},0,0)$&
		$(0,\sqrt{3},0)$
		\\
		$\vec{\sigma}\cdot\vec{\epsilon}^{\dagger}_{s_3}$ &
		$(\sqrt{3},0)$ &
		$(0,\sqrt{3},0)$&
		$(\sqrt{3},0)$&
		$(\sqrt{3},0)$&
		$(\sqrt{3},0,0)$&
		$(0,\sqrt{3},0)$
		\\
		$S(\hat{r},\vec{\sigma},\vec{\epsilon}_{s_1})$  &
		$(0,-\sqrt{6})$ &
		$(\sqrt{3},0,-\sqrt{3})$&
		$(0,\sqrt{6/7})$&
		$(0,-\sqrt{6})$&
		$(0,\sqrt{3/5},-3\sqrt{3/5})$&
		$(3\sqrt{2/5},0,-2\sqrt{3/5})$
		\\
		$S(\hat{r},\vec{\sigma},\vec{\epsilon}^{\dagger}_{s_3})$  &
		$(0,-\sqrt{6})$ &
		$(\sqrt{3},0,-\sqrt{3})$&
		$(0,\sqrt{6/7})$&
		$(0,-\sqrt{6})$&
		$(0,\sqrt{3/5},-3\sqrt{3/5})$&
		$(3\sqrt{2/5},0,-2\sqrt{3/5})$
		\\
		$\vec{\sigma}\cdot\vec{L}$ &
		$0$ &
		$-3$&
		$2$&
		$-2$&
		$1$&
		$-4$
		\\
			$\vec{\epsilon}_{s_1}\cdot\vec{L}$ &
		$(0,0)$ &
		$(0,-\sqrt{3},-\sqrt{3})$&
		$(\dfrac{2\sqrt{3}}{3},-\sqrt{\dfrac{14}{3}})$&
		$(-2\sqrt{\dfrac{1}{3}},-2\sqrt{\dfrac{1}{6}})$&
		$(\sqrt{\dfrac{1}{3}},-\sqrt{\dfrac{5}{3}})$&
		$(0,-\dfrac{4\sqrt{3}}{3},-\dfrac{2\sqrt{15}}{3})$
		\\
			$\vec{\epsilon}^{\dagger}_{s_3}\cdot\vec{L}$ &
		$(0,0)$ &
		$(0,-\sqrt{3},-\sqrt{3})$&
		$(\dfrac{2\sqrt{3}}{3},-\sqrt{\dfrac{14}{3}})$&
		$(-2\sqrt{\dfrac{1}{3}},-2\sqrt{\dfrac{1}{6}})$&
		$(\sqrt{\dfrac{1}{3}},-\sqrt{\dfrac{5}{3}})$&
		$(0,-\dfrac{4\sqrt{3}}{3},-\dfrac{2\sqrt{15}}{3})$
		\\
			$\vec{\epsilon}_{s_1}\cdot\vec{\epsilon}^{\dagger}_{s_3}$	&$\left(\begin{array}{cc}
				1        & 0   \\
			0  &  1          \\
			\end{array}\right) $
			& $\left(
			\begin{array}{ccc}
			1 &  0   &   0    \\
				0 &  1 &   0    \\
				0  &  0  &   1    \\
			\end{array}
			\right)$
			& $\left(
			\begin{array}{cc}
			1          & 0   \\
			0  & 1\\
			\end{array}
			\right)$
			&$\left(
			\begin{array}{cc}
				1        & 0  \\
				0  &  1          \\
			\end{array}
			\right)$
			&$\left(
			\begin{array}{ccc}
			1                   & 0  &   0 \\
			0  &  1         &  0           \\
				0 &  0        &  1    \\
			\end{array}
			\right)$
			&$\left(
			\begin{array}{ccc}
			1        & 0      &  1    \\
			0      &  1                &   0    \\
			0  & 0 &  1     \\
			\end{array}
			\right)$ \\
				$\vec{\epsilon}_{s_1}\cdot{}\vec{\epsilon}^{\dagger}_{s_3}\vec{\sigma}\cdot\vec{L}$	&$\left(\begin{array}{cc}
				0        & 0   \\
				0  &  -3          \\
			\end{array}\right) $
			& $\left(
			\begin{array}{ccc}
				0 &  0   &   0    \\
				0 &  1 &   -2    \\
				0  &  -2  &   -2    \\
			\end{array}
			\right)$
			& $\left(
			\begin{array}{cc}
				-\frac{2}{3}          & -\frac{2}{3} \sqrt{14}   \\
			-\frac{2}{3} \sqrt{14}  &-\frac{1}{3}\\
			\end{array}
			\right)$
			&$\left(
			\begin{array}{cc}
			\frac{2}{3}        & -\frac{2}{3} \sqrt{2}  \\
			-\frac{2}{3} \sqrt{2}  &  -\frac{5}{3}          \\
			\end{array}
			\right)$
			&$\left(
			\begin{array}{ccc}
				-\frac{1}{3}                    & 	-\frac{2}{3}\sqrt{5}   &   0 \\
				-\frac{2}{3}\sqrt{5}  & 	-\frac{2}{3}         &  0           \\
				0 &  0        &  -4    \\
			\end{array}
			\right)$
			&$\left(
			\begin{array}{ccc}
				1        & 0      &  0   \\
				0      &  \dfrac{4}{3}                &  - \dfrac{4}{3}\sqrt{5}    \\
				0  & - \dfrac{4}{3}\sqrt{5}  &  -\dfrac{7}{3}     \\
			\end{array}
			\right)$ \\
				$i(\vec{\epsilon}_{s_1}\times\vec{\sigma})\cdot(\vec{\epsilon}_{s_3}^\dagger)$&
			$\left(
			\begin{array}{cc}
				2        ~~~&   0        \\
				0       ~~~&  -1         \\
			\end{array}
			\right)$  &
			$\left(
			\begin{array}{ccc}
				-1        &  0         &   0     \\
				0        &  2        &   0     \\
				0        &  0         &   -1     \\
			\end{array}
			\right)$ &
			$\left(
			\begin{array}{cc}
				2        ~~~&   0        \\
				0       ~~~&   -1         \\
			\end{array}
			\right)$&
			$\left(
			\begin{array}{cc}
				2        ~~~&   0        \\
				0       ~~~&   -1         \\
			\end{array}
			\right)$&
			$\left(
			\begin{array}{ccc}
				2       &  0         &   0     \\
				0        &  -1         &   0     \\
				0        &  0         &  -1     \\
			\end{array}
			\right)$&
			$\left(
			\begin{array}{ccc}
				-1        &  0         &   0     \\
				0        &  2        &   0     \\
				0        &  0         &   -1     \\
			\end{array}
			\right)$
			\\
				$i(\vec{\epsilon}_{s_3}^\dagger\times\vec{\sigma})\cdot\vec{\epsilon}_{s_1}$&
			$\left(
			\begin{array}{cc}
				-2        ~~~&   0        \\
				0       ~~~&   1         \\
			\end{array}
			\right)$  &
			$\left(
			\begin{array}{ccc}
				1        &  0         &   0     \\
				0        &  -2        &   0     \\
				0        &  0         &   1     \\
			\end{array}
			\right)$ &
			$\left(
			\begin{array}{cc}
				-2        ~~~&   0        \\
				0       ~~~&   1         \\
			\end{array}
			\right)$&
			$\left(
			\begin{array}{cc}
				-2        ~~~&   0        \\
				0       ~~~&   1         \\
			\end{array}
			\right)$&
			$\left(
			\begin{array}{ccc}
				-2       &  0         &   0     \\
				0        &  1         &   0     \\
				0        &  0         &   1     \\
			\end{array}
			\right)$&
			$\left(
			\begin{array}{ccc}
				1        &  0         &   0     \\
				0        &  -2        &   0     \\
				0        &  0         &   1     \\
			\end{array}
			\right)$
			\\
			$S(\hat{r},i\vec{\epsilon}_{s_1}\times\vec{\sigma},\vec{\epsilon}^{\dagger}_{s_3})$
		&$\left(\begin{array}{cc}
			0          &  \sqrt{2}   \\
			-2\sqrt{2}   &  -1          \\
		\end{array}\right) $
		& $\left(
		\begin{array}{ccc}
			0  &  2   &   1     \\
		-1  &  0   &   1    \\
			1  &  -2  &   0    \\
		\end{array}
		\right)$
		& $\left(
		\begin{array}{cc}
			0           & - \sqrt{2/7}   \\
			2\sqrt{2/7}  &  \frac{5}{7} \\
		\end{array}
		\right)$
		&$\left(
		\begin{array}{cc}
			0          &  \sqrt{2}   \\
			-2\sqrt{2}  &  -1          \\
		\end{array}
		\right)$
		&$\left(
		\begin{array}{ccc}
			0                   &  -\frac{1}{\sqrt{5}}  &   -\frac{3}{\sqrt{5}}    \\
			\frac{2}{\sqrt{5}}  &  \frac{4}{5}         &   \frac{3}{5}            \\
			-\frac{6}{\sqrt{5}} &  \frac{3}{5}         &   -\frac{4}{5}           \\
		\end{array}
		\right)$
		&$\left(
		\begin{array}{ccc}
			-\frac{1}{5}        &  2\sqrt{\frac{6}{5}}        &   \frac{2\sqrt{6}}{5}     \\
		 -\sqrt{\frac{6}{5}}         &  0                   &   \frac{2}{\sqrt{5}}     \\
			\frac{2\sqrt{6}}{5}  &  -\frac{4}{\sqrt{5}} &   \frac{1}{5}             \\
		\end{array}
		\right)$ \\
		$S(\hat{r},i\vec{\epsilon}^{\dagger}_{s_3}\times\vec{\sigma},\vec{\epsilon}_{s_1})$
		&$\left(\begin{array}{cc}
			0          &  2\sqrt{2}   \\
			-\sqrt{2}   &  1          \\
		\end{array}\right) $
		& $\left(
		\begin{array}{ccc}
			0  &  1   &  - 1     \\
			-2  &  0   &   2    \\
			-1  &  -1  &   0    \\
		\end{array}
		\right)$
		& $\left(
		\begin{array}{cc}
			0           & - 2\sqrt{2/7}   \\
			\sqrt{2/7}  & - \frac{5}{7} \\
		\end{array}
		\right)$
		&$\left(
		\begin{array}{cc}
			0          &  2\sqrt{2}   \\
			-\sqrt{2}  &  1          \\
		\end{array}
		\right)$
		&$\left(
		\begin{array}{ccc}
			0                   &  -\frac{2}{\sqrt{5}}  &   -\frac{6}{\sqrt{5}}    \\
			\frac{1}{\sqrt{5}}  &  -\frac{4}{5}         &   -\frac{3}{5}            \\
			-\frac{3}{\sqrt{5}} &  -\frac{3}{5}         &   \frac{4}{5}           \\
		\end{array}
		\right)$
		&$\left(
		\begin{array}{ccc}
			\frac{1}{5}        &  \sqrt{\frac{6}{5}}        &  - \frac{2\sqrt{6}}{5}     \\
			-2\sqrt{\frac{6}{5}}         &  0                   &   \frac{4}{\sqrt{5}}    \\
			-\frac{2\sqrt{6}}{5}  &  -\frac{2}{\sqrt{5}} &   -\frac{1}{5}             \\
		\end{array}
		\right)$ \\
		$S(\hat{r},\vec{\epsilon}_{s_1},\vec{\epsilon}^{\dagger}_{s_3})$
	&$\left(\begin{array}{cc}
		0          &  -\sqrt{2}   \\
		-\sqrt{2}   &  1          \\
	\end{array}\right) $
	& $\left(
	\begin{array}{ccc}
		0  &  1   &  - 1     \\
		1  &  0   &   -1    \\
		-1  &  -1  &   0    \\
	\end{array}
	\right)$
	& $\left(
	\begin{array}{cc}
		0           & \sqrt{2/7}   \\
		\sqrt{2/7}  & - \frac{5}{7} \\
	\end{array}
	\right)$
	&$\left(
	\begin{array}{cc}
		0          &  -\sqrt{2}   \\
		-\sqrt{2}  &  1          \\
	\end{array}
	\right)$
	&$\left(
	\begin{array}{ccc}
		0                   &  \frac{1}{\sqrt{5}}  &   -\frac{3}{\sqrt{5}}    \\
		\frac{1}{\sqrt{5}}  &  -\frac{4}{5}         &   -\frac{3}{5}            \\
		-\frac{3}{\sqrt{5}} &  -\frac{3}{5}         &   \frac{4}{5}           \\
	\end{array}
	\right)$
	&$\left(
	\begin{array}{ccc}
		\frac{1}{5}        &  \sqrt{\frac{6}{5}}        &  - \frac{2\sqrt{6}}{5}     \\
		\sqrt{\frac{6}{5}}         &  0                   &   -\frac{2}{\sqrt{5}}    \\
		-\frac{2\sqrt{6}}{5}  &  -\frac{2}{\sqrt{5}} &   -\frac{1}{5}             \\
	\end{array}
	\right)$ \\
		
		\hline
		\hline
	\end{tabular}
\end{table*}

\section{RESULTS And Discussions}\label{Sec: results}
In this work, we systematically discuss whether the $\bar{B}^{(*)}N$ interactions can be bound together to form a molecular.  The method
that we take is the one-boson exchange model, which requires us first to build the $\bar{B}^{(*)}N\to \bar{B}^{(*)}N$ interaction potentials.
The scattering Feynman diagrams for the $\bar{B}^{(*)}N\to\bar{B}^{(*)}N$ reaction at the tree-level are depicted in Figs.~\ref{cc1},
where the $t$-channel $\pi$, $\eta$, $\sigma$, $\rho$, and $\omega$ meson exchanges are considered.  With the effective potential obtained, we can
study the binding property of the system by solving the non-relativistic Schr\"{o}dinger equation
\begin{align}
-\frac{1}{2\mu}[\nabla_r^2-\frac{L(L+1)}{r^2}]\psi(\vec{r})+V(r)\psi(\vec{r})=E\psi(\vec{r})\label{eq46},
\end{align}
where $\mu=m_Nm_{\bar{B}^{}}/(m_N+m_{\bar{B}^{}})$ being the reduced mass for the discussed systems.  $L=0$ is so-called $S$-wave and
often studied in other works. Readers can find the detailed process to solve the Schr\"{o}dinger equation in Refs.~\cite{Zhao:2014gqa,Zhao:2013ffn},
and we do not discuss it here.

We find a bound state if a stable and negative binding energy $E$ appears.  And the masses of the bound state can be determined by
$m=m_{\bar{B}^{}}+m_N-E$.  However, the binding energy depends very sensitively on the parameter $\Lambda$.  To have a reliable
prediction for the bound state requires a good knowledge of $\Lambda$.  Fortunately, there exist some experimental data that can be
used to constrain the value of the cutoff parameter.  Taking $\Lambda=0.862$ GeV, an excellent description of the deuteron binding
energy can be achieved~\cite{Machleidt:1987hj}.  With the  variation of the cutoff from 0.7 GeV to 1.6 GeV,  there exist $D\bar{D}^{*}$ and $B\bar{B}^{*}$ bound states that can
be associated to the $X(3872)$ and $Z_b(10610)$, respectively~\cite{Zhao:2014gqa}.  In Ref.~\cite{Chen:2015loa}, the $\Sigma_c\bar{D}^{*}$
and $\Sigma^{*}_c\bar{D}^{*}$  molecular assignments for $P_c(4380)$ and $P_c(4450)$  can be well explained with $\Lambda=2.35$ GeV and
$\Lambda=1.77$ GeV, respectively.  Larger range of $4.9-5.1$ GeV and $3.6-5.25$ GeV can be found in Refs.~\cite{Zhao:2014gqa,Chen:2020kco}.
However, they argue such cutoffs may be too large for a loosely bound system.  Considering the values adopted in the above literatures,
we take parameter $\Lambda=0.9-2.35$ GeV in this work.

\subsection{The results for $\bar{B}N$ interaction}
\begin{figure}[h!]
\begin{center}
\includegraphics[bb=70 180 750 560, clip, scale=0.48]{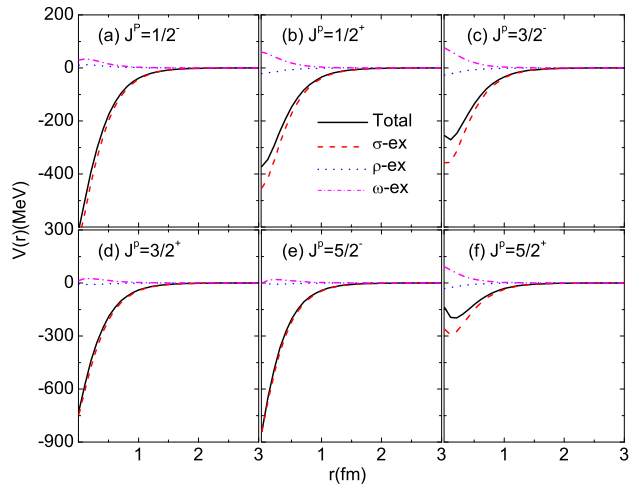}
\caption{The effective potentials of the system $\bar{B}N$ with the parameter $\Lambda=1000$ MeV and different spin-parity assignments.
The solid line represents the total potential. The dashed,  dotted, and dash-dotted lines represent the contributions of the $\sigma$, $\rho$, and $\omega$ exchanges, respectively.}
\label{cc1t}
\end{center}
\end{figure}

In this section, we mainly study whether $\Lambda_b(6146)$ and $\Lambda_b(6152)$ can be explained as $\bar{B}N$ hadronic molecular state.
Since the $\pi{}BB$ and $\eta{}BB$ vertexes are forbidden, there only exist the $\sigma$, $\rho$, and $\omega$ mesons exchange contribution.
Their effective potentials with the parameter $\Lambda=1000$ MeV and different spin-parity assignments  are plotted in Fig.~\ref{cc1t}.
One can find that the $\rho$ and $\omega$ exchanges provide repulsive force while attractive force comes from the $\sigma$ exchange,
which yields a total attractive force.  It is consistent with the interaction of the loosely bound deuteron, in which the repulsive potential
mainly comes from the short-range $\rho$ and $\omega$ mesons exchanges, while the medium-range $\eta$ and $\sigma$
mesons exchange and  the long-range $\pi$ exchange are proved attractive potential.

\begin{table}[h!]
\centering
\caption{Possible bound states for $\bar{B}N$ system with $\Lambda=900-2350$ MeV and different spin-parity assignments (in units of MeV).
$E(-{\cal{F}}_1)$ and $E$ are the eigenvalue without and with the contribution of recoil correction ${\cal{F}}_1$ term, respectively.
Notation $\times$ means no binding solutions.}\label{table3a}
\begin{tabular}{cccc|cccc}\hline\hline
State               ~&$\Lambda$    ~~&$E(-{\cal{F}}_1)$  ~~&$E$            &  State         &$\Lambda$   &$E(-{\cal{F}}_1)$    &$E$              \\ \hline
$J^P=1/2^{-}$       ~&  900        ~~&-20.45             ~~&-16.26         &$J^P=1/2^{+}$   &$900$       &$\times$             &$\times $        \\
		            ~&  950        ~~&-32.68             ~~&-25.45         &                &$\times$    &$\times$             &$\times$         \\
		            ~&  1000       ~~&-46.50             ~~&-35.30         &                &$\times$    &$\times$             &$\times$         \\
                    ~&  1050       ~~&-61.61             ~~&-45.47         &                &$\times$    &$\times$             &$\times$         \\
$\Lambda_b(5912)$   ~&  2791       ~~&                   ~~&-305.61        &                &$ 2350$     &$\times$             &$\times$         \\ \hline
$J^P=3/2^{+}$       ~&  1650       ~~&-19.30             ~~&-3.64          &$J^P=3/2^{-}$   &$900$       &$\times $            &$\times $\\
                    ~&  1700       ~~&-27.76             ~~&-7.80          &                &$\times$    &$\times $            &$\times $        \\
                    ~&  1750       ~~&-37.07             ~~&-12.40         &                &$\times$    &$\times $            &$\times $        \\
                    ~&  1800       ~~&-47.47             ~~&-17.26         &                &$\times$    &$\times $            &$\times $        \\
$\Lambda_b(6146)$   ~&   2205      ~~&                   ~~&-64.74         &                &$2350$      &$\times $            &$\times $        \\ \hline
$J^P=5/2^{+}$       ~&  $900$      ~~& $\times $         ~~& $\times $     &$J^P=5/2^{-}$   &$900$       &$\times $            & $\times $       \\
                    ~&  $\times$   ~~& $\times$          ~~& $\times$      &                &$\times$    &$\times $            &$\times $        \\
                    ~&  $\times$   ~~& $\times$          ~~& $\times$      &                &$\times$    &$\times $            &$\times $        \\
                    ~&  $\times$   ~~& $\times$          ~~& $\times$      &                &$\times$    &$\times $            &$\times $        \\
                    ~&  $2350$     ~~& $\times$          ~~& $\times$      &                &$2350$      &$\times $            &$\times $        \\
\hline\hline
\end{tabular}
\end{table}
For the $^2S_{1/2}$ state, the currently effective potential is enough to obtain the negative and stable eigenvalue $E$.  The numerical
results are shown in Table.~\ref{table3a}.  We find that the $\bar{B}N$  bound state appears when $\Lambda =900$ MeV, and its binding
energy is 16.26 MeV.   With increasing of the cutoff $\Lambda$, the binding energy gets deeper.  It is worth noting that when we increase the
$\Lambda$ to 2791 MeV, far beyond the reasonable range of $900\leq\Lambda\leq2350$ MeV, we obtain a bound state with a mass of about  5912 MeV.
Such result disfavors the assignment of the $\Lambda_b(5912)^0$ as an $S$-wave $\bar{B}N$ molecular state.  It may be a meson-baryon
coupled state including a dominant $\bar{B}N$ component.~\cite{Romanets:2013cqa,Lu:2014ina,Liang:2014eba,GarciaRecio:2012db,Huang:2021ave}.
It is more important that the binding energy of about 305.61 MeV is too large to be a loosely $S$-wave $\bar{B}N$ bound state.  A same $\bar{B}N$
bound state has been found in Ref.~\cite{Carames:2017pfl}.

Now we turn to discuss the possible high wave molecular states.  In Table~\ref{table3a}, we collect the bound properties for the $\bar{B}N$
system with different quantum number configurations.  For the $I(J^P)=0(3/2^{+})$ case, we can obtain bound solutions $E=3.64- 22.45$ MeV
when we tuned cutoff $\Lambda$ from 1650 to 1850 MeV.   If we take an cutoff $\Lambda$ of 2205 MeV, the $\Lambda_b(6146)^0$ can be a good $P$-wave
$\bar{B}N$ molecular candidate.  Although the meson exchanges provide a total attractive potential (see Fig.~\ref{cc1t}), the $\bar{B}N$
bound states with $I(J^P)=0(1/2^{+})$, $I(J^P)=0(3/2^{-})$, and $I(J^P)=0(5/2^{\pm})$ are not survived due to the existence of the strongest
repulsive centrifugal force $L(L+1)/2\mu{}r^2$.  That means the $\Lambda_b(6152)$ state cannot be accommodated in the current $F$-wave
$\bar{B}N$ molecular picture.

\begin{figure}[h!]
\begin{center}
\includegraphics[bb=-50 180 750 660, clip, scale=0.45]{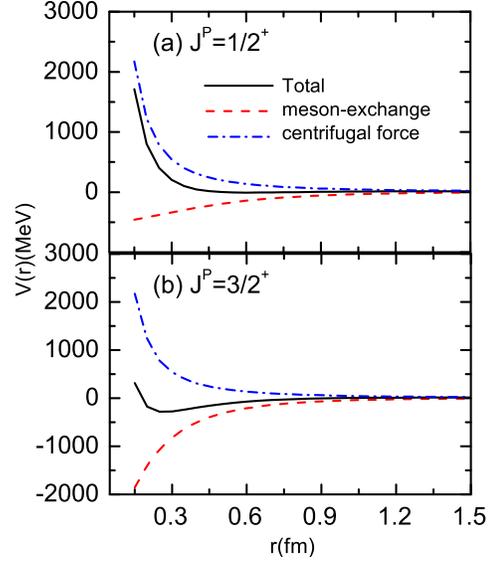}
\caption{The effective potentials of the system $\bar{B}N$ come from the mesons exchange (dash line) and centrifugal force (dash dot line)
contribution at $\Lambda=1700$ MeV. The solid line represents the total potential.}
\label{cc1t2}
\end{center}
\end{figure}

For the angular momentum $L$, the repulsive centrifugal force $L(L+1)/2\mu{}r^2$ of the states $^2P_{1/2}$ and $^2P_{3/2}$ are the same.
However, we find there is a bound state in the $J^P=3/2^{+}$ case, while the molecular in the $J^P=1/ 2^{+}$ case is forbidden.
A possible explanation for this is that the meson-exchange forces in these two cases are quite different.  And we plot these differences
in Fig.~\ref{cc1t2}.  One can find that the meson exchange provides an attraction and a dominant contribution to the $J^P=3/2^{+}$ case,
while the repulsive centrifugal force plays a dominant role in the $J^P=1/2^{+}$ case.  For the same reason as the state $^2P_{1/2}$,
the $D$-wave and the $F$-wave $\bar{B}N$ bound states can be excluded.   And we conclude that if we want to study high waves molecular,
such as the $D$-wave and the $F$-wave, we need a strong attraction exchange potential, which may come from the long-range $\pi$ meson
exchange, the medium-range $\eta$ meson exchange, or gluon exchange. Indeed, in Ref.~\cite{Carames:2017pfl} it was found that the $BN$
system in isospin $I=2$ and spin-parity $J^P=3/2^{-}$ had gluon exchange attractive potential, strong enough to support a deep bound
state.

Obviously, these differences originate from the spin-orbit force contribution for $J^P=1/ 2^{+}$ is two times larger than
that of the $J^P=3/2^{+}$ case (see Table.~\ref{tableqw}).  More important is that in the case of $J^P = 3/2^{+}$ and $J^P =1/2^{+}$,
the spin-orbit force provides an attraction potential and repulsion potential, respectively, in agreement with the conclusion obtained
by using Hund$^{'}$s rule.  These two factors make the meson exchange contribution the most important one in the $J^P=3/2^{+}$ case,
and the bound state can be found.

Now, we discuss the role of recoil correction.   From Table.~\ref{table3a} we find there exists an $S$-wave bound state solution when
the cutoff parameter changes from 900-1100 MeV.  The binding energy without the recoil correction changes from 20.45 MeV to 77.86 MeV,
and the binding energy with the recoil correction is around 16.26-55.76 MeV.  That means the recoil correction decreases the binding
energy and renders this state loosely.  Similar conclusions are obtained for the $P$-wave molecular.  When the binding
energy without the recoil correction is 19.30-58.19 MeV in the range of $\Lambda=1650-1850$ MeV, the binding energy with the recoil
correction can reach 3.64-22.45 MeV.   Hence, we conclude that the recoil correction is necessary to search for a loosely bound state,
and the recoil correction does not alter the conclusion that hadronic molecular states survive.  This is the same as the results by
Zhao et al~\cite{Zhao:2015mga} that the recoil correction is important for the very loosely bound molecular states.
In the following sections, we won't discuss its role again.

\subsection{The results for $\bar{B}N-\bar{B}^{*}N$ interaction}
Since the $\bar{B}N$ channel strongly couples to the $\bar{B}^{*}N$ channel with the scalar and vector meson exchanges,  we study possible
bound states from the $\bar{B}N-\bar{B}^{*}N$ interactions.  The Feynman diagrams for interaction of $\bar{B}N-\bar{B}^{*}N$ are depicted in Fig.~\ref{cc1},
including $\bar{B}N\to{}\bar{B}N$, $\bar{B}N\leftrightarrow{}\bar{B}^{*}N$, and $\bar{B}^{*}N\to\bar{B}^{*}N$ transition diagrams.  For these reactions
we will consider exchanges of scalar mesons $\pi$, $\eta$, $\sigma$ and vector mesons $\rho$, $\omega$.  It is clear that the $\pi$
meson exchange, $\eta/\sigma$ meson exchange, and $\rho/\omega$ meson exchanges contribute to the long, intermediate, and short-range forces,
respectively.  Because there exists the mixing of $S$-wave and $D$-wave or $P$-wave and $F$-wave (see Tab.~\ref{table1-thre}), the $S-D$ mixing
or $P-F$ mixing contributions should be considered.
\begin{table}[h!]
\centering
\caption{Possible bound states from $\bar{B}N$ and $\bar{B}^{*}N$ interaction with $\Lambda=900-2350$ MeV and different spin-parity assignments
(in units of MeV).  $E(-{\cal{F}}_1)$ is the eigenvalue without the contribution of recoil correction ${\cal{F}}_1$ term.  Notation $\times$ means
no binding solutions.  The $\wr$ represent the bound state can be found in a range of $\Lambda$.}\label{table4a}
\begin{tabular}{ccc|ccccc}\hline\hline
State               ~~&$\Lambda$    ~~&$E(-{\cal{F}}_1)$        &~~  State           ~~&$\Lambda$   ~~&$E(-{\cal{F}}_1)$         \\ \hline
$J^P=1/2^{-}$       ~~&  900        ~~&-24.69                   &~~$J^P=1/2^{+}$     ~~&900       ~~&$\times$                  \\
		            ~~&  950        ~~&-39.09                   &~~                  ~~&$\times$    ~~&$\times$                  \\
		            ~~&  1000       ~~&-55.86                   &~~                  ~~&$\times$    ~~&$\times$                  \\
                    ~~&  1050       ~~&-75.06                   &~~                  ~~&$\times$    ~~&$\times$                    \\
$\Lambda_b(5912)$   ~~&  1442       ~~&-336.62                  &~~                  ~~& 2350     ~~&$\times$                  \\ \hline
$J^P=3/2^{+}$       ~~&  1100       ~~&-2.25                    &~~$J^P=3/2^{-}$     ~~& 900       ~~&-2.20                     \\
                    ~~&  1150       ~~&-22.66                   &~~                  ~~& 950       ~~&-5.65                     \\
                    ~~&  1200       ~~&-53.95                   &~~                  ~~& 1000      ~~&-17.77                     \\
$\Lambda_b(6146)$   ~~&  $\wr$      ~~& $\wr$                   &~~                  ~~& 1050      ~~&-40.54                    \\
                    ~~&  1243       ~~&-86.58                   &~~$\Lambda_b(5920)$ ~~& 1381      ~~&-342.58                  \\ \hline
$J^P=5/2^{+}$       ~~&  950        ~~& -2.93                   &$J^P=5/2^{-}$       ~~&$900$       ~~&$\times $                 \\
                    ~~&  1000       ~~& -26.19                  &~~                  ~~&$\times$    ~~&$\times $                 \\
                    ~~&  1050       ~~& -59.95                  &~~                  ~~&$\times$    ~~&$\times $                 \\
$\Lambda_b(6152)$   ~~&  $\wr$      ~~& $\wr$                   &~~                  ~~&$\times$    ~~&$\times $                 \\
                    ~~&  1100       ~~& -101.03                 &~~                  ~~&$2350$      ~~&$\times $                 \\
\hline\hline
\end{tabular}
\end{table}

By solving the Schr\"{o}dinger equation with the inclusion of the coupled-channel effect, the obtained binding energies $E(-{\cal{F}}_1)$ with
the variation of the $\Lambda$ are illustrated in Table.~\ref{table4a}.  We find a bound state with the quantum number $J^P=1/2^{-}$ from the
$\bar{B}N-\bar{B}^{*}N$ interaction is produced.  Its binding energy is between 24.69-75.06 MeV when the cutoff $\Lambda$ lies within $0.90-1.05$
GeV.  It is worth noting that, when we increase $\Lambda$ to about 1.442 GeV, which is a reasonable cutoff, the $\Lambda(5912)$ can be well reproduced.
The best candidate to lodge an $S$-wave $\bar{B}N-\bar{B}^{*}N$ molecule appears in the $J^P=3/2^{-}$ channel.  This is due to the $\Lambda_b(5920)$
can be regarded as meson-baryon molecular, which couple mostly to $\bar{B}^{*}N$ channel~\cite{Romanets:2013cqa,Lu:2014ina,Liang:2014eba,GarciaRecio:2012db,Huang:2021ave}.
In this work, we indeed find the existence of a $J^P=3/2^{-}$ $\bar{B}N-\bar{B}^{*}N$ molecular state that can be associated to the $\Lambda_b(5920)$
due to a reasonable cutoff $\Lambda=1.381$ GeV adopted.

Aside from the $S$-wave molecule, we also want to search for the possible high-wave bound states from the $\bar{B}N-\bar{B}^{*}N$ interaction.
From Table.~\ref{table4a} we find a bound state with $J^P=3/2^{+}$ appears at about $\Lambda=1.200-1.243$ GeV and a bound state with $J^P=5/2^{+}$
is produced with reasonable cutoffs.  Obviously, these two molecular states can be related to the $\Lambda_b(6146)$ and the $\Lambda_b(6152)$,
respectively.

Comparing the results shown in Table.~\ref{table3a} and Table.~\ref{table4a}, it is interesting to find that if the $\Lambda_b(6146)$ is produced as a
pure $\bar{B}N$ bound state, a strong attraction exchange potential is needed due to a larger cutoff corresponds to a stronger interaction.  And the
coupled-channel effect is very important for us to understand the molecular nature of the $\Lambda_b(6152)$.  This is because the interaction between a
$\bar{B}$ meson and a nucleon is not strong enough to form a pure bound state (see Table.~\ref{table3a}).   A possible explanation is that the $\bar{B}N$
component provides a small contribution.  Indeed, it is found from Table.~\ref{table4b} that about $0.99\%-2.72\%$ component of the $\Lambda_b(6152)$ come
from the $\bar{B}N$ channel, while the $\bar{B}^{*}N$ channel is the most important.  However, we found that the $\bar{B}N$ channel proved a larger
contribution to the $\Lambda_b(6146)$, accounting for about $55.99\%-56.58\%$ of the total component.  It leads to an explanation of why the $\Lambda_b(6152)$
can be considered as a pure $\bar{B}N$ bound state and $\bar{B}N-\bar{B}^{*}N$ bound state at the same time.

\begin{table}[h!]
\centering
\caption{Possible bound states from $\bar{B}N$ and $\bar{B}^{*}N$ interaction with $\Lambda=900-2350$ MeV and different spin-parity assignments (in units of MeV).
$E(-{\cal{F}}_1)$ is the eigenvalue without the contribution of recoil correction ${\cal{F}}_1$ term.  Notation $P$ is the probability ($\%$) and the $\wr$ represent
the bound state can be found in a range of $\Lambda$.}\label{table4b}
\begin{tabular}{cccccccc}\hline\hline
State               ~~&$\Lambda$    ~~&$E(-{\cal{F}}_1)$        &~~  $P[\bar{B}N(^{2}P_{3/2})/\bar{B}^{*}N(^{2}P_{3/2}/^{4}P_{3/2}/^{4}F_{3/2})]$   \\ \hline
$J^P=3/2^{+}$       ~~&  1100       ~~&-2.25                    &~~  49.96/29.23/7.09/13.72                               \\
                    ~~&  1150       ~~&-22.66                   &~~  54.46/32.33/6.52/6.69                                \\
                    ~~&  1200       ~~&-53.95                   &~~  55.99/34.04/4.94/5.03                                \\
$\Lambda_b(6146)$   ~~&  $\wr$      ~~&$\wr$                    &~~          $\wr$                                        \\
                    ~~&  1250       ~~&-92.49                   &~~  56.82/34.48/3.78/4.92                                \\ \hline
State               ~~&$\Lambda$    ~~&$E(-{\cal{F}}_1)$        &~~  $P[\bar{B}N(^{2}F_{5/2})/\bar{B}^{*}N(^{4}P_{5/2}/^{2}F_{5/2}/^{4}F_{5/2})]$   \\ \hline
$J^P=5/2^{+}$       ~~&  950        ~~& -2.93                   &~~  5.49/89.26/4.63/0.62                                      \\
                    ~~&  1000       ~~& -26.19                  &~~  2.99/94.03/2.11/0.87                                      \\
                    ~~&  1050       ~~& -59.95                  &~~  2.72/94.51/1.64/1.13                                      \\
$\Lambda_b(6152)$   ~~&  $\wr$      ~~& $\wr$                   &~~         $\wr$                                              \\
                    ~~&  1100       ~~& -101.03                 &~~  0.99/97.96/0.52/0.53                                      \\
\hline\hline
\end{tabular}
\end{table}

\section{summary}
Inspired by the observed $\Lambda_b(6146)$ and $\Lambda_b(6152)$, we study possible molecular states from the interactions of $\bar{B}N$
in a one-boson-exchange approach.  In the framework of the one-boson-exchange model, we have calculated the effective potentials for $\bar{B}N$
system from the $t$-channel $\sigma$, $\rho$, and $\omega$ mesons exchanges.  Then the $\bar{B}N$ bound states
with different quantum number configurations are searched by solving the non-relativistic Schr\"{o}dinger equation.
Since the $\bar{B}N$ channel strongly couples to the $\bar{B}^{*}N$ channel with the scalar and vector meson exchanges,
the coupled-channel effect between these two channels are considered.

Our calculation suggests that recently observed $\Lambda_b(6146)$ cannot only be assigned as a $P$-wave molecular state of $\bar{B}N$
with $I(J^P)=0(3/2^{+})$ but also a $\bar{B}N-\bar{B}^{*}N$ bound state including the mixing of the $P$-wave and $F$-wave.
However, it is not so easy to accommodate the observed $\Lambda_b(6152)$ state as the candidate of $F$-wave $\bar{B}N$ molecular picture.
The $\Lambda_b(6152)$ can be explained as meson-baryon molecular state with a small $\bar{B}N$ component.  The calculation also favors the
existence of two $S$-wave $\bar{B}N-\bar{B}^{*}N$ bound states that can be associated to the $\Lambda_b(5912)$ and $\Lambda_b(5920)$.
The recoil correction is found important for forming a loosely bound state.

\section{Acknowledgments}
This work was supported by the National Natural Science Foundation
of China under Grant No.12104076, the Science and Technology
Research Program of Chongqing Municipal Education Commission
(Grant No. KJQN201800510), and the Opened Fund
of the State Key Laboratory on Integrated Optoelectronics
(GrantNo. IOSKL2017KF19).  Yin Huang acknowledges the YST Program of the APCTP,
the support from the Development and Exchange Platform for
the Theoretic Physics of Southwest Jiaotong University under
Grants No.11947404 and No.12047576,  and the National Natural Science Foundation
of China under Grant No.12005177.

\section{Appendix.A}
In this section, we collect the potential related to the Fig.~\ref{cc1} in the coordinate space.  For the $\bar{B}N \rightarrow \bar{B}N $ reaction, we have
\begin{align}
V_{a/b}(r)_{\rho^0}=&\frac{\beta{}gg_v}{2\sqrt{2}}[Y_1(\Lambda,m_{\rho^0},r)\nonumber\\
                    &+{\cal{V}}_0(m_{\rho^0},\frac{(m_{\overline{B}^0}+2m_p)}{{4m^2_p}m_{\overline{B}^0}},r)],\\
V_{a/b}(r)_{\omega}=&\frac{3\beta gg_v}{2\sqrt{2}}[Y_1(\Lambda,m_{\omega},r)\nonumber\\
                    &+{\cal{V}}_0(m_{\omega},\frac{(m_{\overline{B}^0}+2m_p)}{{4m^2_p}m_{\overline{B}^0}},r)],\\
v_{a/b}(r)_{\sigma}&=-g_{\sigma}g_{\sigma NN} [Y_1(\Lambda,m_{\sigma},r)\nonumber\\
                   &-{\cal{V}}_0(m_{\sigma},\frac{1}{4m^2_P},r)],\\
V_{c/d}(r)_{\rho^{\pm}}&=\frac{\beta{}g g_v}{\sqrt{2}}[Y_1(\Lambda,m_{\rho^{\pm}},r)\nonumber\\
                       &+{\cal{V}}_0(m_{\rho^{\pm}},\frac{(m_{\overline{B}^0}+m_{B^{-}}+4m_p)}{{4m^2_p}(m_{\overline{B}^0}+m_{B^{-}})},r)],
\end{align}
The $\bar{B}N\to \bar{B}^{*}N$ reaction potentials can be shown
\begin{align}
V_{e/f}(r)_\pi&=-\frac{g(D+F)}{2\sqrt{2}ff_\pi}{\cal{V}}_1(m_{\pi^0},r),\\
V_{e/f}(r)_\eta&=\frac{g(D-3F)}{6\sqrt{2}ff_\pi}{\cal{V}}_1(m_{\eta},r),\\
V_{e/f}(r)_{\rho^0}&=\frac{\lambda gg_v}{\sqrt{2}m_p}{\cal{V}}_2(m_{\rho^0},r),\\
V_{e/f}(r)_{\omega}&=\frac{3\lambda gg_v}{\sqrt{2}m_p}{\cal{V}}_2(m_{\omega},r),\\
V_{g/h}(r)_{\pi^{\pm} }&=\frac{g(D+F)}{\sqrt{2}ff_\pi}{\cal{V}}_1(m_{\pi^{\pm}},r),\\
V_{g/h}(r)_{\rho^{\pm}}&=\frac{2\lambda gg_v}{\sqrt{2}m_p}{\cal{V}}_2(m_{\rho^{\pm}},r).
\end{align}

Last, we give the $\bar{B}^{*}N \to \bar{B}^{*}N $ reaction potentials
\begin{align}
V_{i/j}(r)_{\pi^0}&=-\frac{g(D+F)}{2\sqrt{2}ff_\pi}{\cal{V}}_3(m_{\pi^0},r),\\
V_{i/j}(r)_\eta&=\frac{g(D-3F)}{6\sqrt{2}ff_\pi}{\cal{V}}_3(m_{\eta},r),\\
V_{i/j}(r)_{\rho^0}&=-\frac{\beta{}gg_v}{2\sqrt{2}}[(Y_1(\Lambda,m_{\rho^0},r)\nonumber
\end{align}
\begin{align}
                   &+{\cal{V}}_0(m_{\rho^0},\frac{(m_{\overline{B}^{*(0/-)}}+2m_p)}{{4m^2_p}m_{\overline{B}^{*(0/-)}}},r)]\vec{\epsilon}_{\lambda_1}\cdot\vec{\epsilon}_{\lambda_3}^{\dagger}\nonumber\\
                   &+\frac{\lambda{}gg_v}{\sqrt{2}}{\cal{V}}_4(m_{\rho^0},\frac{1}{m_{\overline{B}^{*(0/-)}}},r),\\
V_{i/j}(r)_{\omega}&=-\frac{3\beta{}gg_v}{2\sqrt{2}}[(Y_1(\Lambda,m_{\omega},r)\nonumber\\
                   &+{\cal{V}}_0(m_{\omega},\frac{(m_{\overline{B}^{*(0/-)}}+2m_p)}{{4m^2_p}m_{\overline{B}^{*(0/-)}}},r)]\vec{\epsilon}_{\lambda_1}\cdot\vec{\epsilon}_{\lambda_3}^{\dagger}\nonumber\\
                   &+\frac{3\lambda{}gg_v}{\sqrt{2}}{\cal{V}}_4(m_{\omega},\frac{1}{m_{\overline{B}^{*(0/-),}}},r),\\
V_{i/j}(r)_{\sigma}&=-g_{\sigma}g_{\sigma{}NN}[Y_1(\Lambda,m_{\sigma},r)\nonumber\\
                   &-{\cal{V}}_0(m_{\sigma},\frac{1}{4m^2_P},r)]\vec{\epsilon}_{\lambda_1}\cdot\vec{\epsilon}_{\lambda_3}^{\dagger},\\
V_{k/l}(r)_{\pi^+ }&=\frac{g(D+F)}{\sqrt{2}ff_\pi}{\cal{V}}_3(\pi^{\pm},r),\\
V_{k/l}(r)_{\rho^{\pm}}&=\frac{\beta{}g_1g_v}{\sqrt{2}}[Y_1(\Lambda,m_{\rho^+},r)\nonumber\\
                       &+{\cal{V}}_0(m_{\rho^{\pm}},\frac{(m_{\overline{B}^{*0}}+m_{B^{*-}}+4m_p)}{{4m^2_p}(m_{\overline{B}^{*0}}+m_{B^{*-}})},r)]\vec{\epsilon}_{\lambda_1}\cdot\vec{\epsilon}_{\lambda_3}^{\dagger}\nonumber\\
                        &-\frac{2\lambda{}g_1g_v}{\sqrt{2}}{\cal{V}}_4(m_{\rho^{\pm}},\frac{2}{(m_{B^{*-}}+m_{\overline{B}^{*0}})},r).
\end{align}

In the above, the following equations are used
\begin{align}
{\cal{V}}_0(m,\mu_1,&r)=\mu_1[\frac{1}{4}\nabla_r^2Y_1(\Lambda,m_{\rho^0},r)-\frac{1}{2}\left\{\nabla_r^2,Y_1(\Lambda,m,r)\right\}]\nonumber\\
             &+\frac{1}{16m_{p}^2}\nabla_r^2Y_1(\Lambda,m,r)+\mu_1(\vec{\sigma}\cdot\vec{L}\frac{1}{r}\frac{\partial}{\partial r}Y_1(\Lambda,m,r))\nonumber\\
{\cal{V}}_1(m,r)&=\frac{1}{3}(\vec{\sigma}\cdot\vec{\epsilon}_{\lambda_3}^\dagger)(-\nabla_r^2Y_1(\Lambda,m,r)\nonumber\\
            &+\frac{1}{3}S(\hat{r},\vec{\sigma},\vec{\epsilon}_{\lambda_3}^\dagger)(-r\frac{\partial}{\partial r}\frac{1}{r}\frac{\partial}{\partial r}Y_1(\Lambda,m,r)),\nonumber\\
{\cal{V}}_2(m,r)&=-\vec{\epsilon}_{\lambda_3}^\dagger\cdot\vec{L}(\frac{1}{r}(\frac{\partial}{\partial r}))Y_1(\Lambda,m,r)\nonumber\\
                   &+\frac{1}{3}(\vec{\sigma}\cdot\vec{\epsilon}_{\lambda_3}^\dagger)(-\nabla_r^2Y_1(\Lambda,m,r))\nonumber\\&-\frac{1}{6}S(\hat{r},\vec{\sigma},\vec{\epsilon}_{\lambda_3}^\dagger)(-r\frac{\partial}{\partial r}\frac{1}{r}\frac{\partial}{\partial r}Y_1(\Lambda,m,r))\nonumber\\
{\cal{V}}_3(m,r)&=\frac{1}{3}(\vec{\sigma}\cdot(-i\vec{\epsilon}_{\lambda_3}^\dagger\times\vec{\epsilon}_{\lambda_1}))(-\nabla_r^2Y_1(\Lambda,m,r))\nonumber\\
                &+\frac{1}{3}S(\hat{r},\vec{\sigma},(-i\vec{\epsilon}_{\lambda_3}^\dagger\times\vec{\epsilon}_{\lambda_1}))(-r\frac{\partial}{\partial r}\frac{1}{r}\frac{\partial}{\partial r}Y_1(\Lambda,m,r)),\nonumber\\
{\cal{V}}_4(m,\mu_2,r)&=\frac{1}{m_p}[\frac{1}{3}i(\vec{\epsilon}_{\lambda_1}\times\vec{\sigma})\cdot(\vec{\epsilon}_{\lambda_3}^\dagger)(-\nabla_r^2Y_1(\Lambda,m,r))\nonumber\\
                &+\frac{1}{3}S(\hat{r},i(\vec{\epsilon}_{\lambda_1}\times\vec{\sigma}),\vec{\epsilon}_{\lambda_3}^\dagger)(-r\frac{\partial}{\partial r}\frac{1}{r}\frac{\partial}{\partial r}Y_1(\Lambda,m,r))\nonumber\\
                &-\frac{1}{3}i(\vec{\epsilon}_{\lambda_3}^\dagger\times\vec{\sigma})\cdot(\vec{\epsilon}_{\lambda_1})(-\nabla_r^2Y_1(\Lambda,m,r))\nonumber\\
                &-\frac{1}{3}S(\hat{r},i(\vec{\epsilon}_{\lambda_3}^\dagger\times\vec{\sigma}),\vec{\epsilon}_{\lambda_1})(-r\frac{\partial}{\partial r}\frac{1}{r}\frac{\partial}{\partial r}Y_1(\Lambda,m,r))]\nonumber\\
                &-\mu_2(\frac{1}{3}(\vec{\epsilon}_{\lambda_1}\cdot\vec{\epsilon}_{\lambda_3}^\dagger)(-\nabla_r^2Y_1(\Lambda,m,r))\nonumber
\end{align}
\begin{align}
                &+\frac{1}{3}S(\hat{r},\vec{\epsilon}_{\lambda_1},\vec{\epsilon}_{\lambda_3}^\dagger)(-r\frac{\partial}{\partial r}\frac{1}{r}\frac{\partial}{\partial r}Y_1(\Lambda,m,r)))
\end{align}

\end{document}